\documentstyle[12pt]{article}
\def\beq{\begin{equation}}
\def\eeq{\end{equation}}
\def\bea{\begin{eqnarray}}
\def\eea{\end{eqnarray}}
\def\bq{\begin{quote}}
\def\eq{\end{quote}}

\def\NP{{\it Nucl.Phys.} }
\def\PL{{\it Phys.Lett.} }
\def\PR{{\it Phys.Rev.} }

\def\gappeq{\mathrel{\rlap {\raise.5ex\hbox{$>$}}
{\lower.5ex\hbox{$\sim$}}}}

\def\lappeq{\mathrel{\rlap{\raise.5ex\hbox{$<$}}
{\lower.5ex\hbox{$\sim$}}}}

\begin{document}
\pagestyle{empty}
\begin{flushright}
{CERN-TH.7367/94}
\end{flushright}
\vspace*{5mm}
\begin{center}
{\bf SOME BEAUTIFUL EQUATIONS OF MATHEMATICAL PHYSICS} \\
\vspace*{1cm}
{\bf Daniel Z. Freedman} \\
\vspace{0.3cm}
Theoretical Physics Division, CERN \\
CH - 1211 Geneva 23 \\
and \\
Department of Mathematics and \\
Center for Theoretical Physics \\
M.I.T., Cambridge, MA 02139, U.S.A. \\
\vspace*{2cm}
{\bf ABSTRACT} \\
\end{center}
\vspace*{5mm}
\noindent
The basic ideas and the important role of gauge principles in modern
elementary particle physics are outlined. There are three theoretically
consistent gauge principles in quantum field theory: the spin-1 gauge
principle of electromagnetism and the standard model, the spin-2 gauge
principle of general relativity, and the spin-3/2 gauge principle of
supergravity.
\vspace*{3cm}

\begin{center}
{\it Dirac Lecture} \\
{\it delivered at the} \\
{\it International Centre for Theoretical Physics}\\
{\it Trieste, 19 November 1993}
\end{center}
\vspace*{1cm}
\begin{flushleft}
 CERN-TH.7367/94 \\
July 1994
\end{flushleft}
\vfill\eject

\setcounter{page}{1}
\pagestyle{plain}

    Many quotations   remind us of Dirac's ideas about the
 beauty of
fundamental physical laws. For example, on a blackboard at the
University of Moscow where
visitors are asked to write a short statement for posterity, Dirac
wrote: ``A physical
law must possess mathematical beauty." Elsewhere he wrote: ``A great
deal of my work is
just playing with equations and seeing what they give.". And finally
there is the
famous statement: ``It is more important for our equations to be
beautiful than
to have them fit experiment." This last statement is more extreme than I
can
accept. Nevertheless, as theoretical physicists we have been
privileged to encounter in our education and in our research equations
which
have simplicity and beauty and also the power to describe the real
world. It
is this privilege that  makes scientific life worth living, and it
is this and its close association with Dirac that suggested the title
for this
talk.

  Yet it is a title which requires some qualification at the start.
First,
 I deliberately chose to write ``SOME beautiful equations ..." in full
knowledge that it is only a small subset of such equations that I will
discuss,
chosen because of my own particular experiences. Other theorists could
well
choose an equally valid and interesting subset. In fact it is not a bad
idea
that every theorist ``d'un certain \^age" be required to give a lecture
with the
same title. This would be more creative and palatable than the
alternative
suggestion which is that every theorist be required to renew his/her
professional license by retaking the Ph.D. qualifying exams.

  Second, I do not wish to be held accountable for any precise
definition of
terms such as mathematical beauty, simplicity, naturalness, etc. I use
these
terms in a completely subjective way which is a product of the way I
have
looked at physics for the nearly 30 years of my professional life. I
believe
that equations speak louder than words, and that equations bring
feelings for
which the words above are roughly appropriate.

   Finally, I  want to dispel the notion that I have chosen a
presentation for my own evil purposes. Some listeners probably
anticipate that
they will see equations from the work of Dirac, Einstein and other
true giants. The equations of supergravity will then appear, and the
audience
will be left to draw its own conclusions. I assure you that I have no
such delusions of grandeur. My career has been a mix of good years and
bad
years. If the good years teach good physics, then the bad years teach
humility. Both are valuable.

  The technical theme of this talk is that the ideas of spin, symmetry,
and
gauge symmetry, in particular determine the field equations of
elementary
particles. There are only three gauge principles which are theoretically
consistent.
The first of these is the spin-1 gauge principle which is part of
Maxwell's equations
and the heart and soul of the standard model. The second is the spin-2
gauge principle
as embodied in general relativity. Both theories are confirmed by
experiment. Between
these is the now largely known theoretical structure of supersymmetry
and the associated
spin-3/2 gauge principle of supergravity. Does Nature know about this?
Here, you can
draw your own conclusions.

  This viewpoint is what led me to work on supergravity in 1976. It is
view of the unification of forces before the unification program was
profoundly affected by string theory. However, I confess that I myself
think
far less about unification now than I used to. Instead I think and worry
about
the survival of our profession and our quest to understand the laws of
elementary
particle physics. I hope that it is not a delusion to think that this
presentation may
contribute in a small positive way to the survival of that quest.

    Let us start with the general idea that a particle is a unit of
matter of definite
mass $m$ and spin s. There are two classes of particles, the bosons with
integer
spin 0,1,2 $\ldots$ and the fermions with half-integer spin 1/2, ~3/2
$\ldots$.
We now know that whether a particle is ``elementary" is not an absolute
question. It depends on whether the experiments used to probe it can
achieve a
small enough spatial scale to detect an internal structure of smaller
units. It
is in this way that we have been led in the 20th century from atoms to
nuclei to
quarks. I will simplify that issue by saying that a particle is
elementary if
one can associate with it a wave equation and a local interaction
Lagrangian and
use these to account for experimental results within a certain range of
scales.
Those wave equations are restricted by Lorentz invariance and other
symmetries.
Underlying this is the beautiful mathematical structure which I will
outline.

     A spin-0 particle is described by a real scalar field. If
massless it satisfies a very simple wave equation,
\beq
\Box \phi \equiv \left({\partial^2\over\partial t^2} - \nabla^2\right)
\phi~ (x,t) = 0
\label{1}
\eeq
which is the equation D'Alembert invented to describe acoustic waves in
1747.
If it has a mass then there is another term, and one has the
Klein-Gordon
equation from the 1920's
\beq
(\Box + m^2)\phi = 0~.
\label{2}
\eeq

The particle physics of this equation is also very simple. The equation
is second
order in time. As initial data one must specify both $\phi(x,0)$ and
$\partial_t
\phi(x,0)$ at $t=0$. These two pieces of classical initial data
correspond to a single
quantum degree of freedom; for each possible momentum $p$, there is a
one-particle
state, usually denoted by the ``ket" $\vert p>$, in a fertile notation
we owe to
Dirac.

   Now we come to one of Dirac's major achievements, the wave equation
he
invented in 1927 to describe the spin-1/2 electron in a way consistent
with the
 laws
of special relativity. He postulated a first-order equation for a
four-component complex field
$\psi_\alpha(x,t)$. The equation requires a set of four matrices, now
called $\gamma$
matrices, $\gamma^{\mu}$, satisfying the anti-commutation relations
\beq
\{\gamma^{\mu},\gamma^{\nu}\} = 2\eta^{\mu\nu}~,
\label{3}
\eeq
where $\eta^{\mu\nu} = (+,-,-,-)$ is the Minkowski metric of space-time.
The Dirac equation can then be written in the massless and massive cases
as
\bea
i\partial\llap{$/$} \psi \equiv i\gamma^{\mu}~{\partial\over\partial
x^{\mu}}~\psi(x,t) &=& 0 \nonumber \\
(i\partial\llap{$/$} - m)\psi &=& 0~.
\label{4}
\eea

It would require too long a digression to tell the full story of the
physics
contained in this equation, and I will just list a few things:
\begin{itemize}
\item[1.] an accurate account of the spectrum of hydrogen;
\item[2.] prediction of the magnetic moment of the electron;
\item[3.] negative-energy states and anti-particles;
\item[4.] when applied to other spin-1/2 particles, namely the muon,
proton and
  neutron, the Dirac equation and the system of $\gamma$-matrices
provided
   the framework which established the form of the weak interactions in
   a very exciting chapter of 20th century physics;
\item[5.] the equation is one of the foundations of today's standard
model of particle
  physics. It describes quarks, electrons, muons, and neutrinos, and
their
   strong, electromagnetic, and weak interactions.
\end{itemize}

Despite this broad physical scope the basic particle physics of the
Dirac equation is
straightforward. It is a first-order equation so one must specify the
four
components of $\psi(x,0)$ as initial data. There are four quantum
degrees of
freedom, namely for each momentum $p$, a particle and antiparticle, each
with two
possible spins:
$\vert p,\pm 1/2>$ and $\vert\bar p,\pm 1/2>$.
This is really the same ratio, namely 2/1, of independent classical data
to
particle states, because the four complex components of $\psi$  contain
eight pieces
of real information.

   Following this approach one might think that a spin-1 particle
should be
described by a vector field $A_{\mu}(x,t)$ and the wave equation
\beq
{\rm  massless} \quad\quad\quad \Box  A_{\mu} = 0
\label{5}
\eeq
\beq
{\rm massive}\quad (\Box -m^2)A_{\mu} = 0~.
\label{6}
\eeq
However trouble looms because there is a mismatch between the eight
independent
data for the classical initial value problem and the particle count
required by Poincar\'e invariance, namely two particle states of
helicity $\pm$ 1
in the massless case and three states of helicity $\pm$ 1,0 in the
massive case.
Things get even worse because the extra components of the vector field
give a
quantum theory with negative probabilities, hence unacceptable.

It is here that the principle of gauge invariance comes to the rescue,
with important
consequences both for the linear wave equations of free field theory and
the nonlinear
equations which describe interactions. Gauge invariance is the idea that
part of the
information contained in the field $A_{\mu}$ is unphysical and
unmeasurable, yet it is
difficult and ill-advised to remove it entirely. It is a bit like
writing a triangle on
a piece of paper. The essential information about the triangle is
contained in just
three numbers, the side lengths, but for many purposes, such as to
describe its
relation to another figure on the paper it is useful to introduce a
coordinate system
and specify the coordinates $(x_1,y_1), (x_2,y_2)$ and $(x_3,y_3)$ of
the three vertices.

What is postulated is that the physical information in $A_{\mu}$ is
specified by its
``curl"
\beq
F_{\mu\nu} = \partial_{\mu}A_{\nu} - \partial_{\nu}A_{\mu}~,
\label{7}
\eeq
and this information is unchanged if  $A_{\mu}$ is changed by the
``gradient" of an
arbitrary scalar function $\theta (x)$, viz.
\beq
A_{\mu} \rightarrow A^{\prime}_{\mu} = A_{\mu} + \partial_{\mu}\theta
\label{8}
\eeq
This is called a gauge transformation of $A_{\mu}$. The simplest wave
equation which is invariant under this gauge transformation is
\bea
\partial^{\mu} F_{\mu\nu} &=& 0 \nonumber \\
\partial^{\mu}(\partial_{\mu}A_{\nu} - \partial_{\nu}A_{\mu}) &=& 0
\nonumber \\
\Box A_{\mu} - \partial_{\mu}\partial \cdot A &=& 0~.
\label{9}
\eea
These are all equivalent forms, and the last form shows  that the new
equation
differs from the na\"\i ve one (\ref{4}) by the fairly simple second
term. Yet
 this change is sufficient to solve previous problems, resulting in
\begin{itemize}
\item[1.] a classical initial value problem with four independent
initial data (this
is usually shown using a gauge-fixing procedure  not discussed here);
\item[2.] the quantum theory contains two polarization states $\vert
p,\pm 1>$ of a
massless spin-1 particle; and the gauge property can be used to show
that these states
transform properly under Lorentz transformations;
\item[3.] probabilities are positive.
\end{itemize}

We introduced gauge invariance to describe the photon, but there is a
new and richer
aspect, related to symmetry properties of the Dirac field. Let us look
at the massive
Dirac equation
\beq
i\partial\llap{$/$}\psi - m\psi = 0~.
\label{10}
\eeq
We make a transformation to a new spinor variable
\beq
\psi_{\alpha}(x) \rightarrow \psi^{\prime}_{\alpha}(x) =
e^{i\theta}\psi_{\alpha}(x)~,
\label{11}
\eeq
which is just a change of the complex phase of $\psi (x)$. It is obvious
that
\beq
(i\partial\llap{$/$} - m)\psi^{\prime}(x) =
e^{i\theta}(i\partial\llap{$/$} - m)\psi(x)
= 0~,
\label{12}
\eeq
so $\psi^{\prime}(x)$ satisfies the Dirac equation if $\psi (x)$ does.

This is a symmetry -- a transformation of a set of fields which takes
one solution of
the field equations into another. The phase angle $\theta$ is called the
symmetry
parameter. In this case we have a global or rigid symmetry because
$\theta$ is a
constant, independent of $\vec x$.

However, we are reminded, if only for alphabetic reasons, of our
description of the
electromagnetic field. There we saw that the gauge transformation
(\ref{8}) with an arbitrary function $\theta(x)$ is a symmetry. What
happens if we try to
generalize the previous phase symmetry to
\beq
\psi_{\alpha}(x) \rightarrow \psi^\prime_\alpha (x) = e^{i\theta
(x)}\psi_\alpha (x)~?
\label{14}
\eeq
We must again test whether $\psi^\prime(x)$ satisfies the same field
equation,
and we find
\beq
(i\partial\llap{$/$} - m) \psi^\prime (x) = e^{i\theta (x)}(
i\partial\llap{$/$} - \gamma^{\mu}\partial_\mu\theta - m) ~\psi (x) =
-e^{i\theta}\gamma^\mu \partial_\mu \theta\psi~.
\label{15}
\eeq
So symmetry fails unless $\partial_\mu\theta (x) = 0$, and we are back
to a
global symmetry.

Now comes the powerful step. Suppose that we introduce a new interaction
between $A_\mu (x)$ and $\psi (x)$, using the covariant derivative
\beq
D_\mu\psi \equiv (\partial_\mu - ie A_\mu )\psi (x)
\label{16}
\eeq
and the modified Dirac equation
\beq
i\gamma^\mu D_\mu\psi - m\psi = 0~.
\label{17}
\eeq
It is easy to see that this equation is invariant under the simultaneous
transformation
\bea
\psi(x) \rightarrow\psi^\prime(x) &=& e^{i\theta (x)}\psi(x) \nonumber
\\
A_\mu(x)\rightarrow A^\prime_\mu &=& A_\mu (x) +
i\partial_\mu\theta(x)~.
\label{18}
\eea
So we now have a nonlinear field equation with local symmetry.

The final step is to require that the combined Dirac and Maxwell
equations be
obtained from a gauge invariant Lagrangian, which turns out to be
\beq
{\cal L} = -{1\over 4} F^2_{\mu\nu} + i\bar\psi (\gamma^\mu D_\mu -
m)\psi
\label{19}
\eeq
The $\delta\bar\psi$ variation of ${\cal L}$ produces the
gauge-invariant Dirac
equation, while the $\delta A_\nu$ variation produces the modified
Maxwell
equation
\beq
\partial^\mu(\partial_\mu A_\nu - \partial_\nu A_\mu) =
e\bar\psi\gamma_\nu\psi
\label{20}
\eeq
in which the ``current" $J_\nu = e\bar\psi\gamma_\nu\psi$ is the source.
If we
take the divergence of both sides of the equation, then the left side
vanishes
identically because of $\mu\nu$ antisymmetry, so $J_\nu$ must satisfy
the
equation of continuity
\beq
\partial^\nu J_\nu = {\partial\over\partial t} J_0 - \vec\nabla\cdot\vec
J = 0~.
\label{21}
\eeq
In turn, one can verify that this current conservation equation is
satisfied
because of (\ref{17}) and its complex conjugate. So gauge invariance
produces a
system of field equations linked by subtle consistency conditions. Of
course
one must not forget to mention that what we have obtained in this way
are the
field equations of quantum electrodynamics, which have been verified
experimentally with high precision. Indeed it is this theory and its
coupling
constant $e^2/4\pi\hbar c = 1/137$ that controls, in Dirac's words,
``all of
chemistry and much of physics."

It is worth summarizing what we have done because the same strategy has
worked
at least twice more in this century:
\begin{itemize}
\item[1.] we promoted  the rigid phase symmetry of $\psi (x)$ to a local
symmetry by coupling to the gauge field $A_\mu (x)$ using covariant
derivatives;
\item[2.] in the resulting gauge invariant theory, the conserved current
of the
matter field $\psi$ becomes a source of the gauge field;
\item[3.] a fundamental force of Nature is described in this way.
\end{itemize}

Let us introduce an aesthetic subtheme in this talk, namely the
occurence of
the equations of physics in public art and design. A millenium ago, 1964
to be
exact, I was a postdoctoral fellow at Imperial College in London. I
noticed
then, and on subsequent visits, the frieze over the main door of the
physics
building, where some important equations and facts are carved in black
marble.
I was lucky enough to get (with the considerable help of Dr. K. Stelle
of
Imperial College), some transparencies showing this frieze. There is a
full
view showing four blocks of mathematical material interspersed with
graphics.
And there is an enlargement of the mathematical blocks. The third block
is
devoted to  electromagnetism, with Maxwell's equations in full 19th
century
form very prominent. In the first block there is quantum mechanics with
the
Dirac equation in the upper-right corner. The second block is a mix of
special
relativity, Newtonian gravity (why not general relativity?) and
thermodynamics.
I call your attention only to the numerical relation
\beq
{e^2\over Gm_em_P} = 2.27 \times 10^{39}
\label{22}
\eeq
which gives the ratio of strength of electric and gravitational forces
between
the electron and proton. From this one can easily compute that if there
were no
electromagnetism, the Bohr radius of gravitationally bound hydrogen
would be
$10^{32}$ cm $\sim 10^{15}$ light years. Reciprocally one can see that
it is
only on an energy scale of $10^{19}$ GeV that quantum gravitational
effects
among elementary particles become important.

Because of its importance in the modern picture of particle
interactions, I
must describe the non-Abelian generalization of gauge theory obtained by
Yang
and Mills in 1954. The
mathematical background is a Lie
group $G$ of dimension $N$, with the matrices $T^a_{\phantom{a}ij}$   of
an
$n$-dimensional irreducible representation, structure constants
$f^{abc}$, and commutators
\beq
[T^a,T^a]  = if^{abc}T^c~.
\label{23}
\eeq
At the global level one has $N$ symmetry parameters,  $\theta^a$,
a set of $n$
fermion fields $\psi_{i\alpha}(x)$    and infinitesimal transformation
rule
\beq
\delta\psi_i = i\theta^aT^a_{\phantom{a}ij}\psi_j~.
\label{24}
\eeq
 To achieve local invariance one needs a set of $N$ gauge potentials
$A^a_\mu(x)$.  It
is then straightforward to ``covariantize" equations for $\psi_i$  using
the non-Abelian
covariant derivative
\beq
D_\mu\psi_i \equiv \partial_\mu\psi_i - igA^a_\mu
T^a_{\phantom{a}ij}\psi_j~.
\label{25}
\eeq
The new feature here is that the gauge field is in part its own source.
This
is reflected in its transformation rule
\beq
\delta A^a_\mu = \partial_\mu\theta^a + g ~f^{abc}A^b_\mu\theta^a
\label{26}
\eeq
in which there is both a gradient term similar to the electromagnetic
case (\ref{8}) plus a
    ``rotation" term which survives for constant   $\theta^a$. The
non-Abelian field
strength is nonlinear,
\beq
F^a_{\mu\nu} = \partial_\mu A^a_\mu - \partial_\nu A^a_\mu +
g~f^{abc}A^b_\mu A^c_\nu~,
\label{27}
\eeq
and so is the Yang-Mills field equation
\beq
D^\mu F^a_{\mu\nu}\equiv\partial^\mu F^a_{\mu\nu} + g~f^{abc} A^b_\mu
F^c_{\mu\nu} =
g \bar\psi_i\gamma_\mu T^a_{\phantom{a}ij}\psi_j~.
\label{28}
\eeq
One can show that   $F^a_{\mu\nu}$ and $D^\mu F^a_{\mu\nu}$ transform
homogeneously,
e.g.,
\beq
\delta F^a_{\mu\nu} = g~f^{abc}F^b_{\mu\nu}\theta^c
\label{29}
\eeq
which means that they are covariant under non-Abelian gauge
transformations.

       Non-Abelian gauge invariance is the fundamental principle
underlying
the standard model of elementary particles, and there is strong
experimental
evidence that this model, with gauge group    $SU(3)\times SU(2)\times
U(1)$,    describes
the strong, electromagnetic, and weak forces.

        Our profession is a difficult one.  To find the right field
equations  is only part of the job.  It is far more difficult to solve
those equations in the context of quantum dynamics where each field
variable
is an operator in Hilbert space. Our knowledge of gauge field dynamics
comes
from a combination of experiment and theoretical insight.  It is
fortunate
in many ways that there is a weak coupling regime in which perturbation
theory is valid, and precise calculations using Feynman diagrams can be
performed.

        The only aspect of this dynamics that I will discuss here is the
question of spontaneous symmetry breaking.  This is the phenomenon
that
when
field equations are invariant under a large transformation group $G$,
only a
subgroup, $H \subset G$, need be realized directly in the mass spectrum
and
scattering amplitudes which would be observed experimentally.  For
example,
realization of the full symmetry group $G$ means that all observed
particles can
be organized in multiplets which are representations of $G$ with the
same mass
for all particles in a given multiplet. If the symmetry is broken, then
only
a subgroup $H$ is realized in this way, but there are other observable
signals
of the larger group $G$.  The situation for broken global symmetry is
covered
by the Goldstone theorem, which states that if G has dimension $N$, and
$H$ has
dimension $M$, then there must be $N-M$ massless scalar particles whose
scattering
amplitudes have characteristic properties at low energies.  For broken
gauge
symmetry, one has instead the Higgs mechanism.  The gauge fields
reorganize
into $M$ massless fields of the subgroup $H$, plus $N-M$ fields  which
appear as
massive spin-1 particles. It is quantum dynamics that must tell us
whether symmetry is
broken or not. This depends on whether wave functions invariant under
$G$ or $H$ have
lower energy.

        It is time for another aesthetic interlude, this time from
Washington D.C.  Near the National Academy of Sciences building, and
completely accessible to the public, is a full size statue of Albert
Einstein. He holds a tablet on which the enduring part of his life's
work is
summarized in these three equations
\bea
 R_{\mu\nu} &-& {1\over 2}~g_{\mu\nu}R = \kappa T_{\mu\nu} \nonumber \\
eV &=& h\nu - A \nonumber \\
E &=& mc^2
\label{30}
\eea
from general relativity, the photo electric effect, and special
relativity.
Underneath the equations is his signature.  This is a powerful artistic
statement, which makes one proud to be a physicist. (I thank my MIT
colleague Prof. A. Toomre for
obtaining slides of this statue for me.)

       I want to discuss general relativity very briefly from the
viewpoint
of gauging spacetime symmetry.  The theoretical principle of special
relativity is that physical field equations should be invariant under
translations and Lorentz transformations of space-time.  These are
transformations between two coordinate systems $x^\mu$ and
$x^{\prime\mu}$ related
by
\beq
x^{\prime\mu} = \Lambda^\mu_\nu x^\nu + a^\mu~.
\label{31}
\eeq
In a
special relativity, this is a global symmetry.  There are four
translation
parameters $a^\mu$,  while  $\Lambda^\mu_\nu$   is a matrix of the group
0(3,1)
containing six parameters which describe the relative angular
orientation and velocities
of the two coordinate systems. There is a great deal to say about the
often
counterintuitive effects of the mixing of space and time in special
relativity, but for the purposes of today's talk, I  speak only about
two
formal consequences:
\begin{itemize}
\item[1.] particles are classified by their mass m and spin s;
technically these
numbers specify a representation of the group;
\item
[2.] there is a conserved symmetric stress tensor  $T^{\mu\nu}$ whose
integrals $P^\mu =
f~d^3xT^{0\mu}$ are the energy and momentum of a system of fields or
particles.
\end{itemize}
For the
electromagnetic field this stress tensor is
$$
T^{\mu\nu} = F^{\mu\rho}F^\nu_\rho - {1\over
4}~\eta^{\mu\nu}(F_{\rho\sigma})^2 ~.
$$
If you look carefully you can find the conservation equation
on the Imperial College frieze.

        The gauging of this space-time symmetry is a fairly complicated
process, but the elements are similar to those of the spin-1 gauge
principle.  I must oversimplify and state that one seeks a set of
equations
which are invariant under general coordinate transformations, in which
two
sets of space-time coordinates $x^\mu$  and $x^{\prime\mu}$  are related
in a completely
arbitrary way:
\bea
x^{\prime\mu} &=& a^\mu(x^\nu) \nonumber \\
&\approx & x^\mu + \xi^\mu(x)~.
\label{32}
\eea
where the last form holds for infinitesimal transformations. The gauge
parameter is the
vector $\xi^\mu(x)$, and
the gauge field is a symmetric tensor  $g_{\mu\nu}(x)$  with the
transformation rule
\bea
\delta g_{\mu\nu}(x) &=& \partial_\mu \xi^\rho g_{\rho\nu} +
\partial_\nu \xi^\rho
g_{\mu\rho} - \xi^\rho \partial_\rho g_{\mu\nu} \nonumber \\
&=& D_\mu \xi_\nu + D_\mu \xi_\nu~.
\label{33}
\eea
In the first line one sees a mix of gradient terms plus a translation
term, indicating that the resulting theory is self-sourced.  This is  a
funny way to say that the gravitational field itself carries energy and
momentum.  In the second line I just want to indicate that things can be
organized into covariant derivatives which also simplify the coupling of
$g_{\mu\nu}$ to matter fields.

Finally the analogue of the field strength  $F_{\mu\nu}$  is the
curvature tensor
$R^\lambda_{\mu\rho\nu}$,
from which one forms the Ricci tensor $R_{\mu\nu} =
R^\lambda_{\mu\lambda\nu}$
and the Riemann scalar $R = g^{\mu\nu} R_{\mu\nu}$.
These are the elements of the Einstein field equation
\beq
R^{\mu\nu} - {1\over 2}~g^{\mu\nu} R = \kappa~T^{\mu\nu}
\label{34}
\eeq
in which the source of the curvature is the energy-momentum tensor of
the
matter fields in the system.

        In a single general lecture one can give neither an adequate
technical account of general relativity nor an adequate discussion of
the
ideas it embodies as a theory of gravity.
 I will make three brief comments.

\begin{itemize}
\item[ 1.]  On the formal side, gauge invariance guarantees that the
particle content
of the field  $g_{\mu\nu}$ is the massless spin-2 graviton with two
helicity states
$\vert\vec k,\pm 2>$, with
positive probabilities and interactions which maintain these properties.

\item[2.]   On the side of ideas is the remarkable fact that
$g_{\mu\nu}(x)$  is the
metric tensor of space-time.  So the theory of gravity is a theory of
space-time
geometry, a fact that has captivated many physicists.

\item[3.]  Best of all is the experimental side.  The theory is right in
the
classical domain. Several subtle effects which distinguish Einstein
gravity
from alternative theories (e.g., Newton's) have been observed.  A fairly
recent example is the accurate measurement of the decay of the orbit of
the
binary pulsar at the rate expected from quadrupole gravitational
radiation.
\end{itemize}

       It is  a straightforward matter to take the standard model and
couple
it to gravity by the procedure I have hinted at above. This is
completely
described in several textbooks.  But one learns little  because the
direct quantum effects of gravity are negligible at the energy of any
conceivable particle accelerator. So for practical purposes one can drop
the
gravitational terms and concentrate on the dynamics of particle physics.
Here there are many interesting unsolved problems, and for the last
three
years I have been working on some of them.

       However, most physicists agree that one must eventually
understand
gravity at the quantum level perhaps only as an intellectual question
(but
perhaps more).  One can be fairly certain that there is ``new physics"
at the
quantum gravity or  Planck scale of $10^{19}$ GeV, because theories
obtained by the
straightforward coupling of matter contain uncontrollable infinities.
They
are non-renormalizable in roughly the same way that the effective Fermi
theory of the weak interactions is unrenormalizable.

       The Weinberg-Salam-Glashow model of the electroweak interactions
was  put forward
in 1968. It contained new ideas and was renormalizable.  It predicted
weak neutral
currents which were found in 1972 at the scale of  accelerator
experiments realizable at
that time. The W and Z bosons which were the key to the modification of
the Fermi theory
were found a decade later with masses just below  100 GeV   which was
close to the
weak scale of  300 GeV   at which the Fermi theory necessarily broke
down.
Analogously one can hope that new ideas about quantum gravity could have
 somewhat indirect consequences well below $10^{19}$ GeV  and perhaps
answer some of the questions left open by the standard model.  These
could
include the following.  Does the         group of the standard model
appear as
a subgroup H (unbroken at the weak scale) of a larger unification group
G?  Are
there some restrictions among the free parameters of the model, most of
them
from the poorly understood sector of non-gauge fermion couplings?  This
is the
pragmatic component of the motivation for supersymmetry and
supergravity and
also string theory.  There is also an aesthetic motivation, namely the
search
for beauty and symmetry in physical laws, which I think would have
pleased
 Dirac.

       Supersymmetry is a symmetry of relativistic field theories
connecting
fields of
different spin.  There are transformation rules containing a spinor
parameter
 which rotate a bosonic field into a fermionic superpartner and vice
versa.
That such a symmetry is theoretically consistent was a surprise because
earlier work,
especially the Coleman-Mandula theorem, had indicated that the
invariance groups
permitted in quantum field theory were limited to the Poincar\'e group
of space-time
symmetries and a Lie group $G$ for internal symmetries  as
described above in connection
with non-Abelian gauge theories. Neither contains spin-changing symmetry
operators.
However, in 1971, Golfand and Liktman \cite{aaa} sought to go beyond the
limitations of the
Coleman-Mandula theorem. They wrote down the algebraic relations of an
extension of the
Poincar\'e algebra containing spinor generators, and an interacting
field theory which is
invariant. The mathematical structure is that of a Lie superalgebra,
which was not
considered in earlier work. In 1972, Volkov and Akulov \cite{bb}
obtained another
invariant field theory with a different and pretty structure; it
described a spontaneously
broken form of supersymmetry. Finally in 1973 Wess and Zumino \cite{cc}
discovered
supersymmetry in four-dimensional field theories by generalizing a
structure found in early work
on superstring theory. Their paper contained the basic supersymmetric
theories, with their
off-shell multiplet structure, and systematic rules for constructing
invariant interacting
Lagrangians. The paper of Wess and Zumino was the springboard for the
work of many physicists who
contributed to the formal and phenomenological development of the
subject.

       Let us look at the example of supersymmetric Yang Mills theory,
first obtained by
Ferrara and Zumino \cite{dd} and Salam and Strathdee \cite{ee},  which
is the simplest
interacting theory where you can see  both that supersymmetry
works and that it has some depth.  It is important to look at an
interacting theory
because there are many possible symmetries of a free theory which are
spurious,
because one cannot introduce interactions.  So I will present the full
theory,
 and a guide to the manipulations  needed to show that it is invariant.

       The fields of the theory are the $N$ gauge bosons $A^a_\mu (x)$
and their
superpartners,
 a set of $N$-gauginos  $\chi^a(x)$    which are Majorana spinors.
A Majorana spinor satisfies a linear condition which means that only
four independent
real functions are required as initial data. It describes a spin-1/2
particle which is its own
anti-particle.
The minimal
Lagrangian
 which is gauge invariant, namely
\beq
{\cal L} = -{1\over 4} (F^a_{\mu\nu})^2 + {i\over 2}~
\bar\chi^a\gamma^\mu ~(D_\mu\chi)^a
\label{35}
\eeq
also possesses global supersymmetry.  It is invariant under the
following
 transformations which mix bosons and fermions
\bea
\delta A^a_\mu &=& i\bar\epsilon~\gamma_\mu \chi^a \nonumber \\
\delta\chi^a &=& \sigma^{\mu\nu}F^a_{\mu\nu}\epsilon
\label{36}
\eea
where
\bea
\sigma^{\mu\nu} &=& {1\over 4}~[\gamma^\mu ,\gamma^\nu ] \nonumber \\
D_\mu \chi^a &=& \partial_\mu \chi^a + g ~f^{abc} A^b_\mu \chi^c
\nonumber \\
F_{\mu\nu} &=& \partial_\mu A^a_\nu -  \partial_\nu A^a_\mu + g~f^{abc}
A^b_\mu A^c_\nu~.
\label{37}
\eea
To show supersymmetry in a simplified way, let us establish the
invariance of the free equations of motion.  We need to show that if
$\chi$  and
 $A_\mu$ satisfy the   Dirac and Maxwell equations
\bea
i\gamma^\mu\partial_\mu\chi &=& 0 \nonumber \\
\partial^\mu F_{\mu\nu} &=& 0
\label{38}
\eea
then so do their variations  $\delta\chi$     and     $\delta A_\mu$.
For the Maxwell
equation, we need
\bea
\delta F_{\mu\nu} &=& \partial_\mu\delta A_\nu - \partial_\nu\delta
A_\mu \nonumber \\
&=& i\bar\epsilon (\gamma_\mu\partial_\nu - \gamma_\nu\partial_\mu
)\chi~.
\label{39}
\eea
Then
\beq
\partial^\mu\delta F_{\mu\nu} = i\bar\epsilon (\partial\llap{$/$}
\partial_\nu\chi -
\gamma_\nu~\Box\chi)~.
\label{40}
\eeq
Both terms vanish separately if  $\partial\llap{$/$}\chi = 0$. The
Dirac equation is a
little more
 involved and more instructive
\beq
\gamma^\mu\partial_\mu\delta\chi =
\gamma^\mu\sigma^{\lambda\rho}\partial_\mu
F_{\lambda\rho}\epsilon~.
\label{41}
\eeq
We substitute the standard Dirac matrix identity
\bea
\gamma^\mu\sigma^{\lambda\rho} &=& {1\over 2} ~[\gamma^\mu ,
\sigma^{\lambda\rho}] +
{1\over 2}~ \{\gamma^\mu , \sigma^{\lambda\rho}\}\nonumber \\
&=& {1\over 2} (\eta^{\mu\lambda}\gamma^\rho -
\eta^{\mu\rho}\gamma^\lambda ) + {1\over
2}~i\epsilon^{\mu\lambda\rho\sigma}\gamma_5\gamma_\sigma~.
\label{42}
\eea
in (\ref{41}) and find
\beq
\gamma^\mu\partial_\mu\delta\chi = {1\over 2}
\{(\eta^{\mu\lambda}\gamma^\rho -
\eta^{\mu\rho}\gamma^\lambda ) \partial_\mu F_{\lambda\rho} +
i\gamma_5\gamma_\sigma
\epsilon^{\mu\lambda\rho\sigma}\partial_\mu F_{\lambda\rho} \}\epsilon
{}~.
\label{43}
\eeq
The first two terms vanish by the Maxwell equation above, and the last
vanishes
if one substitutes  $F_{\lambda\rho} = \partial_\lambda A_\rho -
\partial_\rho
A_\lambda$     and uses the fact that  $\epsilon^{\mu\lambda\rho\sigma}$
is totally antisymmetric.

        In the interacting non-Abelian theory things are a little more
complicated.
 The Ricci and Bianchi identities of Yang-Mills theory are required
\bea
[D_\mu ,D_\nu]\chi^a &=& g~f^{abc} F^b_{\mu\nu}\chi^c \nonumber \\
\epsilon^{\lambda\mu\nu \rho} D_\mu F^a_{\nu\rho} &=& 0
\label{44}
\eea
and one then finds that the term   (in $\delta{\cal L}$)
\beq
g~f^{abc}~\bar\epsilon \gamma^\mu \chi^a \bar\chi^b \gamma_\mu \chi^c
\label{45}
\eeq
must vanish as the final test of invariance.  It can be shown to vanish
as a
consequence of the Fierz rearrangement identity for the $\gamma$
matrices and the
crucial fact that the spinor quantities $\chi^a (x)$  and $\epsilon$
must anticommute
because of
 the Pauli exclusion principle.

        It is worthwhile to summarize the ingredients of the proof:
\begin{itemize}
\item[a.]  the Ricci and Bianchi identities which are fundamental to the
non-Abelian
 gauge invariance;
\item[b.]  properties of the $\gamma$ matrix algebra used in the
relativistic treatment
of spin;
\item[c.]  anti-commutativity of fermionic quantities required by the
connection of
 particle spin and statistics.
\end{itemize}

       If the discussion above does not  convince you that
supersymmetry
is a
principle of great depth, then let me describe one more fact.  This is
the
relation of supersymmetry to the space-time transformations of the
Poincar\'e
group.  For any physical field   $\Phi(x)$    of a supersymmetric
theory one can make
repeated
supersymmetry variations, with spinor parameters,  $\epsilon_1$     and
 $\epsilon_2$.
The commutator
 of two transformations is
\beq
(\delta_{\epsilon_1}\delta_{\epsilon_2} -
\delta_{\epsilon_2}\delta_{\epsilon_1}) \Phi(x)
= i\bar\epsilon_1\gamma^\mu \epsilon_2~~\partial_\mu \Phi(x)~.
\label{46}
\eeq
Thus the commutator is an infinitesimal translation in space-time with
displacement
parameter $\delta a^\mu = i\bar\epsilon_1\gamma^\mu\epsilon_2$.  So
supersymmetry is a
``square root" of translations in
much the same way that the Dirac equation is said to be the ``square
root" of
the scalar wave equation.  This is already enough to see that the local
form of
 supersymmetry must involve gravity, and we will return to this shortly.

       In a theory with a non-Abelian gauge group $G$, we have seen
that the
fields are
organized in representations
of $G$.  In a supersymmetric theory there
is the
analogous Poincar\'e  super-algebra, including translations, Lorentz
and SUSY
transformations.  Fields are organized in multiplets of this algebra,
the basic
 ones contain fields of spins
\vfill\eject
\bea
{\rm chiral~multiplet}~~(1/2,0^+,0^-) &&\nonumber \\
{\rm gauge~multiplet}\quad\quad\quad (1,1/2) && \nonumber \\
(3/2,1) && N \geq 2~{\rm supergravity}\nonumber \\
N = 1~{\rm supergravity} \quad (2,3/2) && \nonumber \\
(5/2,2) && \nonumber \\
(s,s - 1/2) &&
\label{47}
\eea

A field theory with local supersymmetry is called a supergravity theory.
  Let us see what is required for such a theory  by applying
what we have learned about the spin-1 and spin-2 gauge principles.  We
want
invariance with respect to transformations with an arbitrary spinor
function
$\epsilon_{\alpha}(x)$, so we should expect to require a gauge field
with an additional
vector index, a vector-spinor field  $\psi_{\mu\alpha}(x)$. A free
field theory for
$\psi_{\mu\alpha}(x)$ had been
formulated in 1941 by Rarita and Schwinger,  describing a spin-3/2
particle.  For a
Majorana field, their Lagrangian is
\beq
{\cal L} = -{1\over
2}~\epsilon^{\lambda\rho\mu\nu}~
\bar\psi_{\lambda}\gamma_5\gamma_\mu~\partial_\nu\psi_\rho~.
\label{48}
\eeq
 One can easily see that it is invariant under the gauge transformation
$\delta\psi_\rho = \partial_\rho\epsilon$.
So the spin-3/2 field is the natural candidate for the gravitino, the
superpartner of the graviton, and we should expect that the particle
content of
the basic supergravity theory should be given by the (2, 3/2)
supermultiplet
 above.

 However it was not clear that the theory could be mathematically
consistent  because of
the
infamous history of attempts to add interactions to the
Rarita-Schwinger
 theory.  All such attempts had led to inconsistencies.  For example if
 $\psi_\rho$ is coupled to an electromagnetic field using the covariant
derivative,
$D_\nu\psi_\rho = (\partial_\nu - ieA_\nu)\psi_\rho$, the resulting
theory, although
formally relativistic, has
propagation of signals at velocities faster than light.  We now know
that such
problems arise when the interactions fail to incorporate the gauge
invariance of the free theory.

 Our approach \cite{ff} to the construction of supergravity was to
start with the minimal
elements required in a gravitational Lagrangian with fermions.  These
were:
 \bea
{\rm vierbein}~~e^a_\mu && \nonumber \\
{\rm spin~connection}~~\stackrel\circ\omega_{\mu ab} &=& {1\over
2}~[e^\nu_a(\partial_\mu e_{b\nu} - \partial_\nu e_{b\mu}) +
e^\rho_ae^\sigma_b
(\partial_\sigma e_{c\rho}) e^c_\mu - (a\leftrightarrow b)] \nonumber \\
{\rm curvature~tensor}~~R_{\mu\nu ab} &=& \partial_\mu
\stackrel\circ\omega_{\nu ab} +
\stackrel\circ\omega_{\mu a}^{\phantom{\mu a}c}\stackrel\circ\omega_{\nu
cb} -
(\mu\leftrightarrow\nu )  \nonumber \\
{\rm Lorentz~covariant~derivative}~~D_\nu\psi_\rho &=& (\partial_\nu +
{1\over
2}~\stackrel\circ\omega_{\nu ab} \sigma^{ab})\psi_\rho~.
\label{49}
\eea
 From these we formed the  Lagrangian
\bea
{\cal L} &=& {\cal L}_2 + {\cal L}_{3/2} \nonumber \\
&\equiv& -{\det e\over 4\kappa^2}~e^{a\mu}
e^{b\nu}R_{\mu\nu ab} -{1\over 2}
{}~\epsilon^{\lambda\rho\mu\nu} \bar\psi_\lambda
\gamma_5~e^a_\mu \gamma_a D_\nu
\psi_\rho
\label{50}
\eea
where $\kappa$ is the gravitational coupling constant.  The first term
is the standard
pure gravity action in vierbein form, and the second the
Rarita-Schwinger
Lagrangian with minimal gravitational coupling.  This Lagrangian is
acceptable
from the viewpoint of the spin-2 gauge principle, and the next question
is
 whether it is locally supersymmetric.

  For this one needs transformation rules.  It was natural to postulate
\beq
\delta\psi_\rho = {1\over\kappa}~D_\rho\epsilon =
{1\over\kappa}~(\partial_\rho\epsilon + {1\over 2}
\stackrel\circ\omega_{\rho
ab}\sigma^{ab}\epsilon )
\label{51}
\eeq
because this is both gravitationally covariant and contains the expected
mix of
 gradient plus Bose-Fermi mixing terms.  The vierbein variation
\beq
\delta~e^a_\mu = -i~\kappa\bar\epsilon\gamma^a\psi_\mu
\label{52}
\eeq
is almost uniquely determined by invariance arguments.

I will present the first
steps in the proof of local supersymmetry which shows that
terms of order $\kappa^{-1}\bar\epsilon \psi$ vanish in the variation of
the
action. In conventional vierbein gravity the variation of ${\cal L}_2$
for any
$\delta e^a_\mu$ is
\beq
\Delta S_2 = {1\over 2\kappa^2} \int d^4x \det e(R^{a\mu}
- {1\over 2} e^{a\mu}R)
\delta e_{a\mu}~.
\label{53}
\eeq
It is the Einstein tensor in frame form that multiplies $\delta
e^a_\mu$.
We now compute the $\delta\psi$ variation of ${\cal L}_{3/2}$, getting a
factor
of two by varying $\bar\psi_\lambda$ and $\psi_\rho$ according to the
rules for
Majorana spinors,
\bea
\Delta_1 S_{3/2} &=& -{1\over \kappa} \int d^4
x~\epsilon^{\lambda\rho\mu\nu}\bar\psi_\lambda
\gamma_5\gamma_\mu D_\nu D_\rho
\epsilon \nonumber \\
&=& -{1\over 4\kappa} \int
d^4
x~\epsilon^{\lambda\rho\mu\nu}\bar\psi_\lambda
\gamma_5\gamma_\mu R_{\nu\rho bc}
\sigma^{bc} \epsilon
\label{54}
\eea
where we have used the gravitational Ricci identity in the last line. We
now use
(\ref{42}) in the form
\beq
\gamma_\mu\sigma^{bc} = {1\over 2} (e^b_\mu\gamma^c - e^c_\mu\gamma^b) +
{1\over 2} i e^a_\mu \epsilon^{abcd}\gamma_5\gamma_d
\label{55}
\eeq
When this is inserted in (\ref{54}) the first two terms give
contractions
\beq
\epsilon^{\lambda\rho\mu\nu} R_{\nu\rho\mu b} = 0
\label{56}
\eeq
which vanish due to the first Bianchi identity for the curvature tensor.

After use of $\bar\psi_\lambda\gamma_d \epsilon =
-\bar\epsilon\gamma_d\psi_\lambda$, which holds for
anti-commuting Majorana
spinors, we are left with
\beq
\Delta_1 S_{3/2} = {1\over 8\kappa} i \int d^4x
\epsilon^{\mu\lambda\rho\nu}\epsilon^{abcd}
e^a_\mu R_{\nu\rho bc}\bar\epsilon
\gamma_d\psi_\lambda~.
\label{57}
\eeq
It is less fun, but straightforward, to compute the contraction of the
$\epsilon$ tensors
\bea
\epsilon^{\mu\lambda\rho\nu}\epsilon^{abcd}
e^a_{\phantom{a}\mu}R_{\nu\rho bc}
&=&
2\det e(e^{b\lambda}e^{c\rho} e^{dr} + e^{c\lambda} e^{d\rho}e^{b\nu} +
e^{d\lambda}e^{b\rho}e^{c\lambda})R_{\nu\rho bc} \nonumber \\
&=& 4 \det e(R^{d\lambda} - {1\over 2} e^{d\lambda}R)~.
\label{58}
\eea
When this is inserted in (\ref{57}) and (\ref{52}) is inserted in
(\ref{53}) we find an
exact cancellation!

The situation is similar to that of supersymmetric gauge theories. The
cancellation is due to the combined effects of the gravitational Ricci
and
Bianchi identities and the Dirac $\gamma$-matrix algebra. This lowest
order
cancellation showed  that we were on the right track, but  there was
more
work to be done because there is a non-vanishing variation $\Delta_3
S_{3/2}$ of
order $\kappa \bar\epsilon \psi^3$.

Here we struggled for many weeks because it was
hard to perceive a pattern in quantities with so many indices. Finally
we
 devised a systematic approach involving:
\begin{itemize}
\item[a)] a general ansatz for a modified gravitino transformation of
order
$\delta^\prime\psi = \kappa\bar\epsilon\psi^2$;
\item[ b)]  an analogous general ansatz for a contact Lagrangian of
order
${\cal L}_4 = \kappa^2(\bar\psi\psi )^2$.
\end{itemize}
 The  $\delta^\prime\psi$  variation of  ${\cal L}_{3/2}$
 and the  $\delta\psi$  variation
of  ${\cal L}_4$   give additional order $\kappa\epsilon\psi^3$
 terms and we were able to find unique choices for
 $\delta^\prime\psi$ and  ${\cal L}_4$
to make the total
 variation vanish.  Fierz rearrangement was required here.

  Unfortunately new and complicated terms of order
  $\kappa^3\bar\epsilon\psi^5$ are
generated by $\delta e$
and   $\delta^\prime\psi$    variations of ${\cal L}_4$.  One could
show easily that no
further
modification of the framework could be made, and these terms had to
vanish or
the theory failed.  We were able to show that they vanished by a
computer calculation in
FORTRAN language with explicit input of the $\gamma$-matrices and a
program to
 implement the anti-symmetrization implicit for fermionic variables.

 Soon thereafter an important simplification of the resulting theory
was obtained by Deser
and Zumino \cite{ggg}, with a further simplifying step \cite{hh}
somewhat later. This involved the
idea that the gravitino modifies the space-time geometry by including
torsion. The net result is
that the Riemannian spin connection $\stackrel\circ\omega_{\mu ab}$ is
replaced by
\beq
\omega_{\mu ab} = \stackrel\circ\omega_{\mu ab} + {1\over
2}~i\kappa^2~(\bar\psi_\mu\gamma_a\psi_b - \bar\psi_\mu\gamma_b\psi_a -
\bar\psi_a\gamma_\mu\psi_b)
\label{59}
\eeq
in the Lagrangian (\ref{50}) and transformation rule (\ref{51}). This
grouping of terms gives a
complete and succint definition of the theory, and a simpler proof of
invariance.

  The subsequent development of supergravity included:
\begin{itemize}
\item[1.]   coupling supergravity to the chiral and gauge multiplets of
global
supersymmetry; these couplings involve conserved supercurrents and
super-covariant derivatives;
there are now relatively simple tensor methods to
 obtain the most general form of these theories;
\item[ 2.]   developing extended supergravity with $N \leq 8$
gravitinos; the maximal $N =
8$ theory was once thought to be the best candidate for a unified field
theory.
\item[ 3.]  Higher dimensional supergravity culminating in the 10 and 11
dimensional
theories.  The 10 dimensional version is important both historically
and
 practically for superstrings.
\end{itemize}

 Earlier we said that the only theoretically consistent gauge
principles are
those of spin-1, spin-2, and spin-3/2.  This  information comes from a
set of
theorems, due to Coleman and Mandula and Haag, Lopuszanski
and Sohnius,
which
limit the symmetries
permitted in an interacting theory.  These
theorems hold under certain assumptions which must be examined
critically.  But
it appears that they are essentially correct.  For example, one can
write free
field theories for a spin-5/2 field, but attempts to include
interactions have
 all failed.

 There is time to describe only very briefly what now appears to be the
most
plausible scenario for experimental verification of these ideas.  This
is the
global supersymmetric extension of the standard model with fields
grouped in
chiral and gauge multiplets.  The known quark, lepton, gauge, and Higgs
fields
all have  superpartners.  One then couples this large set of matter
fields
to supergravity. Observed supersymmetry requires that a particle and
its
superpartner must have the same mass.  This is decidedly false, so one
must
expect supersymmetry breaking, and superpartners are predicted with
masses
 between 100 GeV and 1 TeV.

 Without the supergravity couplings explicit mechanisms for the
symmetry
breaking have not been found.  There could be a subtle dynamical
breaking
mechanism, but in any case the spontaneous global supersymmetry
breaking would
give a Goldstino, a massless spin-1/2 particle that is excluded
experimentally.  So the role
 of supergravity in these models is to break supersymmetry such that the
gravitino becomes massive by a super-Higgs mechanism, without
generating a
 cosmological constant.

The first model which correctly described this super-Higgs mechanism was
obtained by Polonyi
\cite{jj}. General studies of the conditions for the super-Higgs effect
by Cremmer, Julia, Scherk,
Ferrara, Girardello and van Nieuwenhuizen \cite{kk} and by Cremmer,
Ferrara, Girardello and Van
Proeyen \cite{lll} also contain the most general $N = 1$ supergravity
actions. It is this work
which has been widely applied to supersymmetric extensions of the
standard model. A very early
discussion \cite{mm} of the super-Higgs effect for $N$ spin-3/2 fields
and $N$ Goldstone fermions
is incorrect both for general $N$ and   in the special case $N = 1$.

 Discovery of the superpartners is the key requirement to confirm the
picture
of broken supersymmetry, but there is also a less direct set of
predictions
related to the unification scale of gauge coupling of the standard
model, the
rate of proton decay and the masses expected for the top quark and
Higgs
bosons.  There is now favorable experimental evidence on the
unification scale
and the observed lower limit on the proton lifetime.  These facts
appear quite
naturally in the supergravity models but not in the simplest forms of
theories
 without supersymmetry.

 In this lecture I have not done justice to string theory and the
beautiful
 ideas it contains.  Therefore I must clearly state that the $N = 1$
supersymmetry/supergravity framework I have discussed cannot give a
complete
theory.  There are non-renormalizable infinities which require new
physics at
the Planck scale.  A more fundamental superstring theory could well be
correct
and there are well studied scenarios by which such a theory can lead,
for
energies less than   $10^{19}$~GeV, to an effective $N = 1$
supergravity theory.

  I have not had the time to discuss some of the pragmatic features of
supersymmetry/ supergravity theories which make them attactive as a
candidate
for physics beyond the standard model. There are review articles to
consult
about this very active subject of research.  Instead what I have tried
to say is that these
theories
are based on the only theoretically consistent symmetry principle not
so far
confirmed in Nature.  This suggests that it is historically  inevitable
for
supersymmetry to play a role.  Of course this could be as dangerous as
the
prediction that ``capitalism contains within itself the seeds of its
own
destruction." Experiment is the ultimate test of theoretical
speculation.  New
experiments are needed together with theorists who are willing to
devote a good
 part of their effort to support the experimental enterprise.

\vfill\eject
\noindent
{\bf Note}

We can include explicit reference only to a few of the original papers
on supersymmetry and
supergravity. Many important papers are omitted, and it is fortunate
that they are reprinted and
reviewed in the following collections which are also a very good way to
learn the subject.
\begin{itemize}
\item[a.] ``Supersymmetry and Supergravity", ed. M. Jacob, North
Holland, Amsterdam (1986), a
collection of Physics Reports by J.Ellis, P. Fayet and S. Ferrara, H.
Haber and G. Kane, C.
Llewellyn Smith, D. Nanopoulos, P. van Nieuwenhuizen, H. Nilles, A.
Savoy-Navarro and M. Sohnius.
\item[b.] ``Supersymmetry", 2 Vols., ed. S. Ferrara, North Holland,
Amsterdam (1987).
\item[c.] ``Supergravities in Diverse Dimensions", 2 Vols., eds. A.
Salam and E. Sezgin, North
Holland/World Scientific (1989).
\end{itemize}

\end{document}